Présentation d'une méthodologie d'évaluation de la faisabilité de la substitution sensorielle par électrostimulation linguale pour la prévention des escarres chez le blessé médullaire paraplégique.

Dr Alexandre Moreau-Gaudry[1,2,4], Dr Anne Prince[3], Mr Olivier Chenu[4], Pr Jacques Demongeot[1,4], Mr Yohan Payan[4]

[1]Centre d'Innovation Technologique, [2]Centre d'Investigation Clinique, Centre Hospitalier et Universitaire, 38 706 La Tronche Cedex, FRANCE
[3]Centre Médico Universitaire Daniel Douady, 38660 St Hilaire du Touvet, FRANCE
[4]Laboratoire TIMC, Equipe Gestes Médico-Chirurgicaux Assistés par Ordinateur, Institut d'Ingénierie et de l'Information en Santé, 38706 La Tronche Cedex, France.
Tel : 04 76 76 92 60. Emel : AMoreau-Gaudry@chu-grenoble.fr

**Résumé** - Cet article rapporte les évolutions technologiques et ergonomiques entreprises en vue de l'acceptation dans le quotidien du lésé médullaire d'un nouveau dispositif de prévention des escarres. Ce dispositif repose sur le principe de la substitution sensorielle qui consiste à suppléer à une modalité sensorielle déficiente (la sensibilité fessière du paraplégique), une modalité sensorielle fonctionnelle (sa sensibilité linguale). Cet article présente brièvement le protocole expérimental de la recherche biomédicale prospective, monocentrique, randomisée, contrôlée, en groupe parallèle, équilibré et ouverte relative à l'évaluation de la faisabilité de l'utilisation par des paraplégiques de ce nouveau dispositif.

# 1 Introduction

## 1.1 La suppléance sensorielle

Paul Bach-y-Rita et ses collaborateurs [1,2] ont montré que des stimuli caractéristiques d'une modalité sensorielle déficiente (la vision des aveugles) pouvaient être « remplacés » par des stimuli d'une autre modalité sensorielle fonctionnelle (leur touché) grâce à l'utilisation d'un système de suppléance visuo-tactile. Leurs travaux pour améliorer ce dispositif ont conduit à choisir la langue comme organe idéal de substitution sensorielle du fait de sa forte capacité discriminative et de la grande conductivité de la salive qui permet la mise en œuvre d'une technologie basée sur une électro-stimulation à très faible énergie.

## 1.2 Objectif

Le projet de recherche vise à compenser les déficits sensoriels des paraplégiques par l'utilisation d'un dispositif d'électrostimulation linguale (ensemble d'électrodes placées en bouche) couplé à des capteurs de pression inclus au sein d'une nappe de pression, dans le but de réduire l'incidence des escarres de pression chez le paraplégique au niveau de la région fessière. Après une première évaluation de la faisabilité d'une telle approche chez des sujets sains sans lésion médullaire [3], des améliorations ergonomiques ont été entreprises en vue de faciliter l'acceptation du nouveau dispositif médical par le lésé médullaire dans son quotidien. L'évaluation de la faisabilité de cette approche et de son acceptabilité chez le paraplégique est réalisée dans le cadre d'une recherche biomédicale avec dispositif médical.

## 2. Matériels
### 2.1 La nappe de pression

Développée par la société canadienne Vista Medical® et mise à notre disposition dans le cadre de ce projet, la nappe de pression se présente sous la forme d'une mince feuille (figure 1) qui est disposée sur le fauteuil du paraplégique. Dans l'application développée, elle est connectée à un ordinateur portable et permet l'acquisition des forces de pression exercées sur la nappe en 32x32 points et à une fréquence de 5 Hz. Avant chaque acquisition journalière, la nappe est calibrée à l'aide d'un dispositif pneumatique développé par cette société.

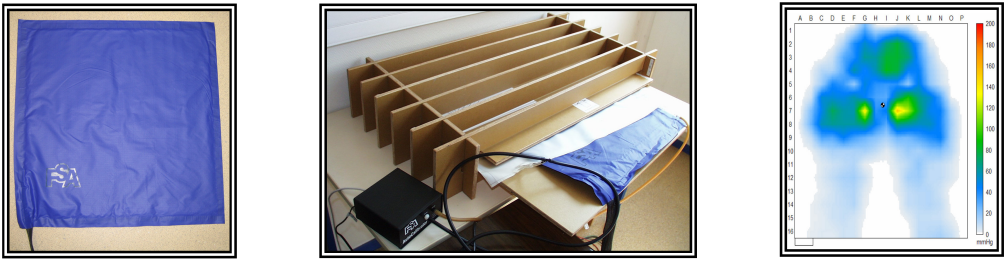

Fig. 1 : Présentation de la nappe de pression FSA de la société Vista Medical® (figure de gauche). Illustration du dispositif de calibration qui permet de calibrer la nappe par mise en pression à des étalons déterminés (figure du milieu). Illustration des pressions exercées par un sujet en position assise sur la nappe, avec hyperpression (zones de couleur jaune et rouge) au niveau des régions d'appui ischiatiques (figure de droite).

### 2.2 Le « Tongue Dispay Unit » TDU

L'ergonomie du dispositif initialement développé [3] a été améliorée en collaboration avec les sociétés Coronis® et Guglielmi Technologies Dentaires®. La miniaturisation de l'ensemble de l'électronique du système d'électrostimulation linguale permet son incorporation dans une orthèse palatine (figure 2). Ce système communique avec l'ordinateur portable par radiofréquence selon un protocole fiable et sécurisé. L'orthèse palatine laisse affleurer à sa surface inférieure les 36 électrodes tactiles (6x6). Ces électrodes sont au contact de la langue. Lorsqu'une électrode est activée, le sujet ressent, à ce niveau, un « picotement » à la surface de sa langue. Le réglage du niveau de la stimulation électrique permet de s'adapter à la perception de l'individu.

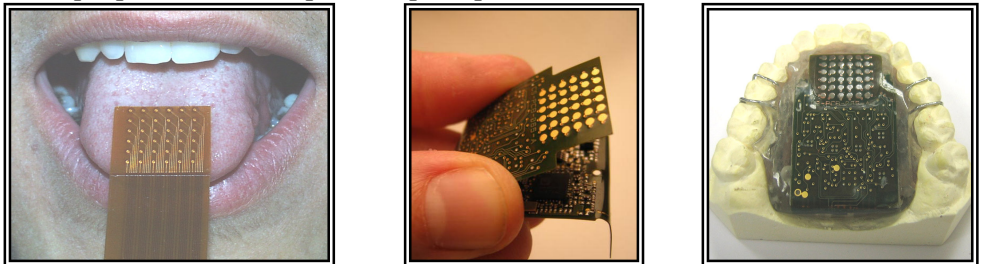

Fig 2 : La première version du Tongue Display Unit était filaire et incompatible avec une utilisation quotidienne (figure de gauche). Suite à un processus de miniaturisation (figure du centre), l'ensemble de l'électronique est noyée dans une orthèse palatine (figure de droite), qui laisse affleurer les 36 électrodes de stimulation linguale à sa surface inférieure. La communication du dispositif avec

l'ordinateur portable est assurée par voie hertzienne selon un protocole fiable et sécurisé.

## 3 Une recherche biomédicale

### 3.1 Le cadre législatif

L'évaluation de ce nouveau dispositif médical fait l'objet d'un protocole expérimental dans le cadre législatif imposé par le code de la santé publique relatif aux recherches biomédicales. Le promoteur de cette étude est le Centre Hospitalier et Universitaire de Grenoble. Le Comité de Protection des Personnes a donné un avis favorable. Cette recherche est en cours de déclaration auprès de l'Agence Française de Sécurité Sanitaire des Produits de Santé. Elle se déroulera dans le respect de la Loi Informatique et Liberté et des Bonnes Pratiques Cliniques.

### 3.2 Les objectifs

L'objectif principal de cette recherche est d'évaluer si le nouveau dispositif médical développé permet l'amélioration, de manière déterministe et adaptée, de la distribution spatiale au cours du temps de la pression exercée sur les tissus mous, à l'interface saillies-osseuse/fauteuil : le sujet paraplégique peut-il se déplacer « en pression » et de manière adaptée suite à une information de direction de déplacement (du buste) stimulée électriquement sur la langue de manière périodique (toutes les deux minutes) (figure 3)? La direction de déplacement stimulée est calculée de manière automatique par l'ordinateur par analyse des cartes de pression, de manière à ce que les zones de surpression préalablement identifiées soient soulagées suite au déplacement du buste du sujet dans la direction électrostimulée.

Les principaux objectifs secondaires portent sur l'évaluation de manière qualitative et quantitative de l'étape de calibration linguale, étape nécessaire à l'utilisation du dispositif du fait de l'anisotropie spatiale individuelle de la sensibilité et sur l'évaluation de manière qualitative du ressenti du sujet paraplégique suite à l'utilisation de ce nouveau dispositif médical.

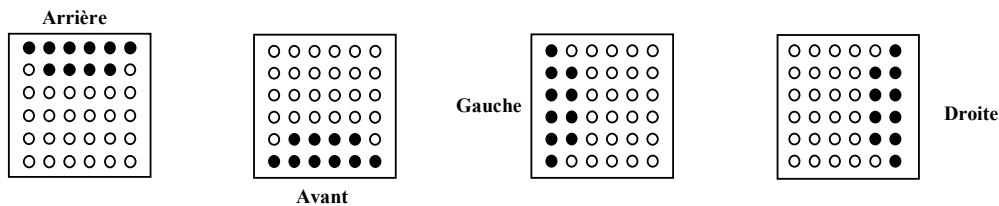

Fig 3 : Chaque sujet équipé du TDU doit s'efforcer de déplacer son buste vers l'arrière, l'avant, la gauche et la droite suite à l'activation des 10 électrodes respectivement postérieures, antérieures, gauches et droites.

### 3.3 Le déroulement pratique

Un schéma synoptique est présenté figure 4. Cette étude, monocentrique, se déroulera au Centre d'Investigation Clinique du CHU de Grenoble (CHUG) en collaboration avec le Centre d'Innovation Technologique de Grenoble et le Centre Médico Universitaire Daniel Douady. Le calcul de l'effectif nécessaire à la réalisation de cette recherche biomédicale a conduit à 24 sujets paraplégiques. Ils seront inclus dans cette étude prospective après information et visite médicale préalables à la recherche biomédicale, vérification des critères d'inclusion et d'exclusion, et signature du consentement de participation. Suite à une étape de randomisation équilibrée, deux groupes de sujets seront constitués : ceux sans (groupe A) et avec (groupe B) utilisation du dispositif médical endobuccal de substitution sensorielle. Chaque sujet du groupe A réalisera, à une semaine d'intervalle, au cours d'une même activité standardisée, un enregistrement d'une durée de 60 minutes des pressions exercées à l'interface saillies-osseuses/fauteuil. Chaque sujet du groupe B, suite à la prise de ses empreintes dentaires nécessaires à la confection de l'orthèse palatine, réalisera les mêmes épreuves cliniques que précédemment, le dispositif de substitution sensorielle étant activé lors de la seconde épreuve. C'est au cours de cette épreuve que sera évalué, de manière qualitative, le ressenti des sujets par rapport au dispositif médical.

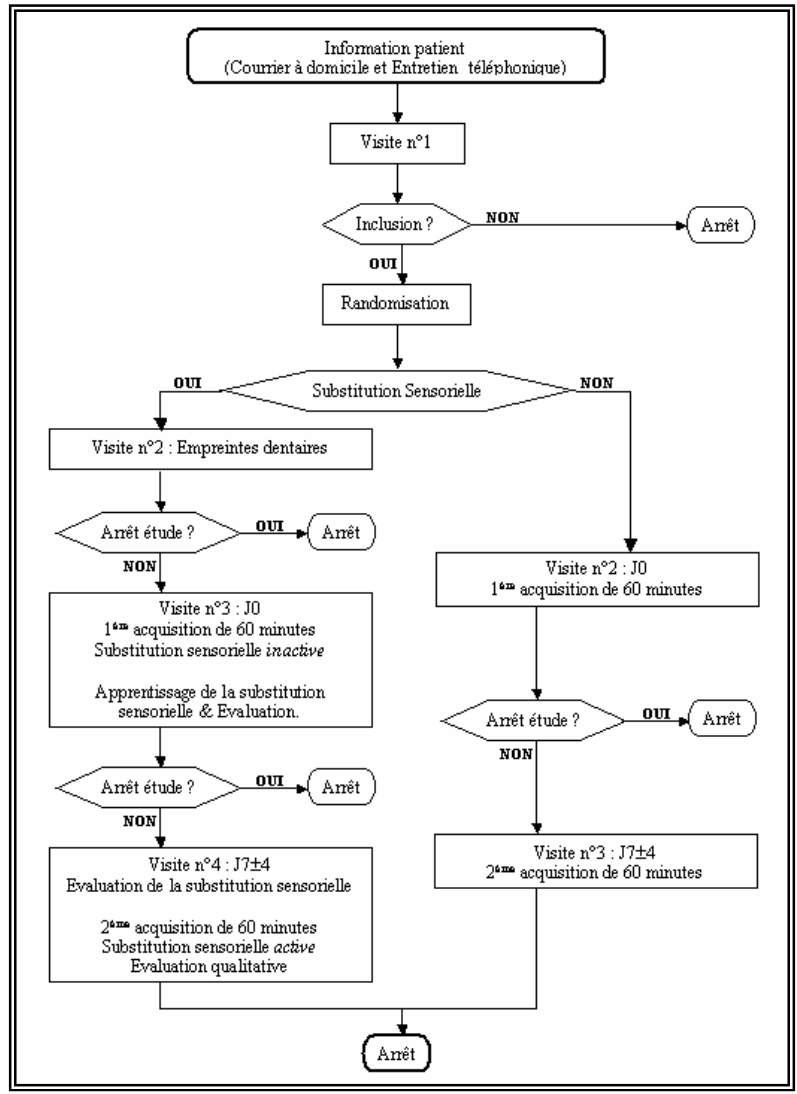

Fig 4 : Schéma synoptique de l'organisation de cette recherche biomédicale.

### 3.4 L'analyse statistique

Le monitorage de l'étude permettra l'obtention de données de qualité. Ces données seront analysées en intention de traiter. L'unité statistique est la différence, par sujet, des réponses adaptées en pression entre les deux épreuves cliniques. L'analyse des données quantitatives de l'objectif principal portera sur la comparaison, entre les deux groupes A et B, des moyennes/médianes des unités statistiques. Les tests statistiques utilisés seront adaptés à la distribution de l'ensemble des unités statistiques. Les objectifs secondaires seront analysés principalement de manière descriptive.

**4 Au total**

Les probants premiers résultats de l'étude préliminaire, sur sujets sains, de l'utilisation de la substitution sensorielle à la « prévention » des escarres nous ont amené à concevoir et développer un nouveau dispositif médical en vue de son acceptation dans le quotidien du sujet paraplégique. Grâce aux progrès technologiques, le système électronique du TDU, qui communique par voie hertzienne avec l'ordinateur portable de l'application, est maintenant totalement intégré dans une orthèse palatine. Cette nouvelle technologie nécessite désormais d'être évaluée cliniquement avec une méthodologie de qualité en vue d'estimer, à terme, de manière exhaustive et objective, le service médical effectivement rendu au blessé médullaire paraplégique pour prévenir la formation d'une escarre.

Références
[1] P. Bach-y-Rita, C. Collins, F. Saunders, B. White and L. Scadden, "Vision Substitution by Tactile Image Projection", *Nature*, vol. 221, pp. 963-964, March 1969.
[2] P. Bach-y-Rita, "Late postacute neurological rehabilitation: neuroscience, engineering and clinical programs", *Archives of Physical Medicine and Rehabilitation*, vol. 84, no. 8, pp. 1100-8, August 2003.
[3] A. Moreau-Gaudry, F. Robineau, P.F. André, A. Prince, P. Pauget, J. Demongeot Y. Payan. « Utilisation de la substitution sensorielle par électro-stimulation linguale pour la prévention des escarres chez les paraplégiques. Etude préliminaire ». *L'escarre*, vol. 30, pp. 34-37, Juin 2006.